\documentclass[twocolumn,sort&compress,noshowpacs,preprintnumbers,amsmath,amssymb,aps,floatfix,prl]{revtex4}
 
\usepackage{graphicx}
\usepackage{epstopdf}
\usepackage{amsmath}
\usepackage{multirow}
\usepackage{times}
\usepackage{amssymb}
\usepackage{dcolumn}%
\usepackage{bm}%
\usepackage{upgreek}
\usepackage{wasysym}
\usepackage{verbatim}
\usepackage{blindtext}
\usepackage[driverfallback=dvipdfm]{hyperref}

\hypersetup
{
	colorlinks=true,
	linkcolor=blue,
	citecolor=blue
}

\begin{document}

\title{Boson peak decouples from elasticity in glasses with low connectivity}

\author{A.\@ Giuntoli}
\affiliation{Dipartimento di Fisica ``Enrico Fermi'', 
Universit\`a di Pisa, Largo B.\@Pontecorvo 3, I-56127 Pisa, Italy}

\author{D.\@ Leporini}
\email{dino.leporini@unipi.it}
\affiliation{Dipartimento di Fisica ``Enrico Fermi'', 
Universit\`a di Pisa, Largo B.\@Pontecorvo 3, I-56127 Pisa, Italy}
\affiliation{IPCF-CNR, UOS Pisa, Italy}

\begin{abstract}
\noindent
We perform molecular-dynamics simulations of the vibrational and the elasto-plastic properties of polymeric glasses and crystals and  corresponding atomic systems. We evidence that the elastic scaling of the density of states in the low-frequency boson peak (BP) region is different in crystals and glasses.
Also, we see that the BP of the polymeric glass is nearly coincident with the one of the atomic glasses, thus revealing that the former - differently from elasticity - is controlled by non-bonding interactions only. Our results suggest that the interpretation of the BP in terms of macroscopic elasticity, discussed in highly connected systems, does not hold for systems with low connectivity.
\end{abstract}


\maketitle

\textbf{\textit{Introduction.-}}
Anomalies in the thermodynamics of glasses with respect to the crystals are observed in specific heat and 
thermal conductivity \cite{BinderKob}. There is general consensus that the difference has to be ascribed to the low-frequency portion of the distribution $g(\omega)$ of the frequencies
of the vibrational states (DOS). In particular, an excess of vibrational states over the level predicted by the Debye squared-frequency law
of the long-wavelength acoustic waves has been universally noted, thus resulting in a so called boson peak (BP) when plotting the reduced DOS $g(\omega)/ \omega^2 $ \cite{BinderKob}.
Models for the BP dealt with quasi-local vibrational states due to soft anharmonic potentials \cite{BuchenauLocalModeBPPRB91,ParshinSchoberLocalizedBPPRB03}, local inversion-symmetry breaking \cite{ZacconeBP_PRB16},
phonon-saddle transition in the energy landscape \cite{ParisiBPNature03}, elastic heterogeneities
\cite{SchirmacherDiezemannBPElasticHetPRL98,GotzeBPElasticHetPRE00,ElliottBPVanHovePRL01,BarratElasticHetBosonPeakNonAffinePRL06,RuoccoSchirmacherBPElasticHeterogSciRep13}
and broadening and shift of the lowest van Hove singularity in the corresponding reference crystal \cite{ChumakovBPVanHovePRL11} due to the distribution of force constants
\cite{ShengDOSSpringScience91,SchirmacherDiezemannBPElasticHetPRL98,ElliottBPVanHovePRL01}.

The work described here is motivated by the observation that the BP frequency window corresponds to wavelengths where the homogeneous picture assumed by the Debye model of elastic bodies become questionable. Therefore,
there is considerable interest in understanding if the BP region retains information included in the macroscopic elasticity and in the DOS of the relevant crystal with both positive 
\cite{SchirmacherDiezemannBPElasticHetPRL98,ElliottBPVanHovePRL01,MonacoBPElasticPRL06,FontanaCaponiBPElasticitySilicaPRB07,TanakaBP_FreeVol_NatMat08,MonacoFiorettoBP_polymerResinPRL09,ChumakovBPVanHovePRL11} and negative \cite{SokolovBPpolymJCP97, SokolovMWvsFragilMM04,AndrikopoulosBPElasticityJNCS06,NissSokolovBPPolymPRL07,SokolovSimionescoBPPolymerPRB08,RuoccoSchirmacherBPElasticHeterogSciRep13,RamosBP_ElasticPRL14} evidences. The latter often found in polymers by changing temperature \cite{SokolovBPpolymJCP97}, molecular weight \cite{SokolovMWvsFragilMM04} and pressure \cite{NissSokolovBPPolymPRL07,SokolovSimionescoBPPolymerPRB08}. 

Reconsideration of earlier numerical studies \cite{ShengDOSSpringScience91} suggests that the coupling between BP and elasticity is driven by high microscopic connectivity. Prompted by that remark,
we carry out molecular-dynamics (MD) simulations of the vibrational and the elasto-plastic properties of systems with low or no connectivity to scrutinize this controversial coupling. More specifically, we compare {\it atomic glasses} to  {\it glasses of linear polymers} where experiments evidenced decoupling between BP and macroscopic elasticity \cite{SokolovBPpolymJCP97,SokolovMWvsFragilMM04,NissSokolovBPPolymPRL07,SokolovSimionescoBPPolymerPRB08}. Glasses are compared to their crystalline counterparts and all the solids are chosen with nearly matched density, local and global order.

\textbf{\textit{Key findings.-}} We show that the BP of the glasses under study is poorly coupled to the macroscopic elasticity and question
the presence of a universal connection between vibrational dynamics and elasticity moving from crystal to glass. 
We also show that the BP of the atomic and polymeric glasses {\it coincide} even if their moduli are rather {\it different}. The finding shows that  the BP of our polymer model
is {\it unaffected} by connectivity and under {\it exclusive control} of non-bonded weak interactions. This offers an explanation of  the larger sensitivity of the polymeric BP
to pressure than elasticity \cite{NissSokolovBPPolymPRL07,SokolovSimionescoBPPolymerPRB08} (since non-bonded interactions are expectedly much more anharmonic than the bonded ones), and is consistent with
the noted role of the softer regions \cite{TanakaBP_FreeVol_NatMat08} and the weak interactions  \cite{TorellBPPolymerINTERMolecularPhilMag98,AndrikopoulosBPElasticityJNCS06,RyzhovBP_Intermolec2008, ZacconeBosonPeakPolymerMM18} in BP. 
On this basis, we suggest that the BP of poorly or not connected systems cannot be interpreted as in highly connected systems,
e.g. strong glassformers \cite{MonacoBPElasticPRL06,FontanaCaponiBPElasticitySilicaPRB07,ChumakovBPVanHovePRL11} and network polymeric glasses  \cite{MonacoFiorettoBP_polymerResinPRL09}, where strong coupling between BP and macroscopic elasticity is observed.

\textbf{\textit{Background.-}} The static shear elastic modulus is written as $G = G_A - G_{NA}$ where $G_A$ is the affine modulus and $-G_{NA}$ expresses the negative non-affine correction, i.e. a {\it softening effect}, due to the possible inhomogeneous deformation at molecular length scales 
\cite{Hoover69,BarratAQS13,BarratBosonPeakNonAffinePRB02,BarratPRE09}. $G_A$ quantifies the stiffness of an effective spring accounting for the random oscillations of a tagged particle in the cage of the immobile surrounding ones \cite{Boon}.
Under broad assumptions, the non-affine correction $G_{NA}$ of an athermal isotropic material with unit-mass particles having number density $\rho$
 is related to the DOS (normalized to $1$) via the sum rule \cite{LemaitreMaloneyJStatPhys06,ZacconeNonAffinePRB11}:
\begin{equation}
G_{NA} = 3  \rho  \int_0^{\infty} \frac{g(\omega)}{\omega^2} \, \Gamma_G(\omega) \, d \omega
\label{GLemMal}
\end{equation}
$\Gamma_G(\omega)$ is the correlator of the projections on the normal modes at frequency $\omega$ of the forces acting on the particles which would result from an affine, i.e. homogeneous, displacement of all the particles in the strain direction. Under the action of these forces, the particles displace from the initial affine position to their final non-affine equilibrium position.
  $\Gamma_G$ is small in the presence of limited local asymmetry of particle configurations
 \cite{LemaitreMaloneyJStatPhys06,ZacconeBP_PRB16}, e.g., in Bravais lattices \cite{LemaitreMaloneyJStatPhys06} but not  in non-centro-symmetric and disordered lattices \cite{ZacconeBP_PRB16} where $G$ differs from $G_A$ significantly \cite{BarratBosonPeakNonAffinePRB02,BarratPRE09}. 
 Eq.\ref{GLemMal} shows clearly the coupling between the reduced DOS - and then BP - with the non-affine part of the modulus. The coupling is present also in the extension of Eq.\ref{GLemMal} to account for finite-frequency viscoelasticity \cite{LemaitreMaloneyJStatPhys06,ZacconeSoftMatter18} with parameter-free predictions for polymer glasses at finite temperature \cite{ZacconeSoftMatter18}. However, $G_A$ is not coupled to DOS \cite{Boon,LemaitreMaloneyJStatPhys06,ZacconeNonAffinePRB11,ZacconeBP_PRB16,ZacconeSoftMatter18} so that the link between BP and  the modulus $G$ is not obvious.

\textbf{\textit{Models.-}}
MD simulations were carried out with the LAMMPS code (http://lammps.sandia.gov) \cite{PlimptonLAMMPS} to simulate atomic and polymeric systems of $N=500$ particles.
In the polymer samples, non-bonded particles interact with a truncated Lennard-Jones potential:
$U^{LJ}(r)=\varepsilon  [ \left (\sigma^*/r \right)^{12} - 2\left (\sigma^*/r \right)^6 \,]+U_{cut}$
where $\sigma^*=2^{1/6}\sigma$ is the position of the potential minimum with depth $\varepsilon$. The value of the constant
$U_{cut}$ is chosen to ensure $U^{LJ}(r)=0$ at $r \geq r_c=2.5\,\sigma$.
Polymer chains have $M=10$ monomers per chain and bonded monomers interact with an harmonic potential $U^b(r)=k(r-r_0)^2$ with $k=2500 \, \varepsilon  / \sigma^2$ and $r_0=0.97\,\sigma$.
In the atomic samples only the Lennard-Jones potential $U^{LJ}(r)$ is present.
All quantities are expressed in term of reduced Lennard-Jones units with unit monomer mass and Boltzmann constant.
The limited size of the samples allows isothermal crystallization of the polymer samples and suppresses dislocation-mediated elasto-plasticity in atomic crystals.
We start from an NPT equilibration in the supercooled liquid phase of fifty-six polymeric samples and fifty-eight
atomic samples at comparable densities and let the systems equilibrate.
After several equilibration times, we observe {\it spontaneous} crystallization of several samples and obtain forty-two polymeric crystals and thirty-four atomic crystals.
We then quench all the crystalline systems and the liquid systems just after equilibration to temperature $T=10^{-3}$ and pressure $P=0$ in a single time step $\Delta t=0.003$ 
and, following known protocols \cite{BarratAQS13,NicolaElastico}, later allow them to relax with an NPT run to let the total energy stabilize. 
Liquids quenched with this protocol form glassy solids.
We thus obtain four classes of solids, i.e. polymer crystals, polymer glasses, atomic crystals and atomic glasses, 
with final densities $ \rho \simeq 1.118,  1.075,   1.052 , 1.010$ respectively. The slightly different packings are due to the different connectivity and solid phase. 

\begin{figure}[t]
\begin{center}
\includegraphics[width=\linewidth]{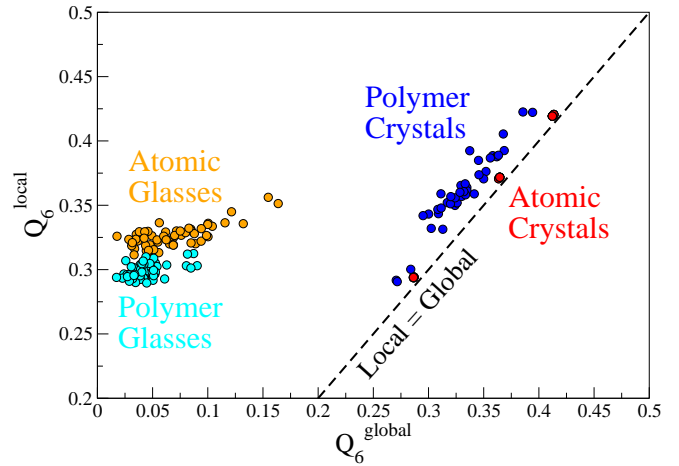}
\end{center}
\caption{Correlation plot of the Steinhardt parameters $Q_6^{local}$ and $Q_6^{global}$ \cite{Steinhardt83} of all the samples under study at $T=10^{-3}$, $P=0$.
For ideal crystals $Q_6^{local} = Q_6^{global}$. The global order is rather small in glasses, while their local order does not differ too much from the crystalline one.}
\label{structureComparison}
\end{figure}

\textbf{\textit{Structural features.-}} We first characterize the structure of the solids under study with the global, $Q_6^{global}$, and local, $Q_6^{local}$,
bond-orientational order parameters defined by Steinhardt \textit{et al.} computed in the first neighbor shell \cite{Steinhardt83,GiuntoliCristallo}. 
In the presence of ideal crystalline order $Q_6^{global}=Q_6^{local}$, whereas for glasses $Q_6^{global} \ll Q_6^{local}$.
Fig.\ref{structureComparison} is a correlation plot  $Q_6^{local}$ vs. $Q_6^{global}$. 
It is seen that: 
i) crystals are nearly ideal, with a structure comparable to a body-centered cubic (Bcc) lattice \cite{GiuntoliCristallo,note}, ii) glasses and crystals have similar local ordering.
The above standard analysis characterizes both the local and the global order of the solids of interest and is enough for the present purposes.
Nonetheless, it is worth noting that in atomic glasses and defective crystals, both with harmonic bonds with identical stiffness, a new  order parameter based on local inversion-symmetry breaking was reported to better correlate with the BP than the bond-orientational local order parameters \cite{ZacconeBP_PRB16}.

The fact that crystals have near Bcc order is of interest. The Bcc DOS has a characteristic low-frequency
van Hove (VH) singularity, ascribed to a transverse [110]-T${}_1$ mode where the particles move into the large octahedral
holes of the lattice, so that the Bcc DOS is richer in soft modes than the close-packed structures \cite{HafnerDOSBCC}.
This feature is appealing to test the interpretation of the BP of the glass as a modified VH singularity of its reference crystal \cite{SchirmacherDiezemannBPElasticHetPRL98,ElliottBPVanHovePRL01,ChumakovBPVanHovePRL11}.

\begin{figure}[t]
\begin{center}
\includegraphics[width=0.95\linewidth]{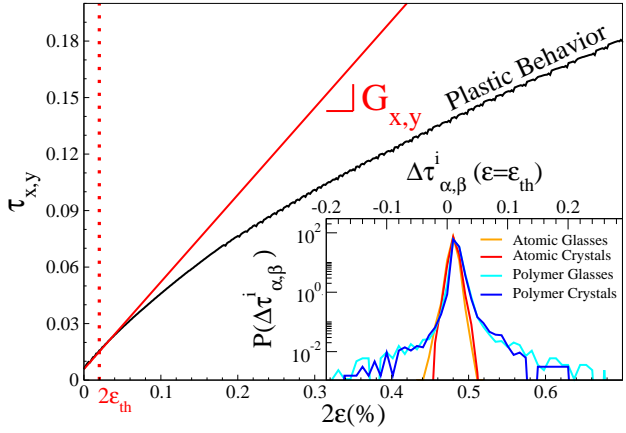}
\end{center}
\caption{Typical stress-strain curve for a single shear plane under athermal quasi-static shear deformation 
(the loaded system is a polymer crystal; the differences with the other systems are small).
The curve is characterized by an initial linear elastic regime followed by a region with tiny, but apparent, sudden stress drops signaling plastic events which become much larger at higher strain,
e.g. see   \cite{NicolaElastico}.
The elastic modulus $G_{x, y}$ is the slope for small deformations, $\varepsilon \le  \varepsilon_{th} =  10^{-4}$. 
Inset: distributions of the stress changes per particle $\Delta \tau^i_{\alpha, \beta}$ at $\varepsilon = \varepsilon_{th}$ for all samples in all the deformation planes.
The weak wing in the region of negative drops signals the presence of some local plastic events even in the macroscopic linear regime.}
\label{carico} 
\end{figure}

\begin{figure}[t]
\begin{center}
\includegraphics[width=\linewidth]{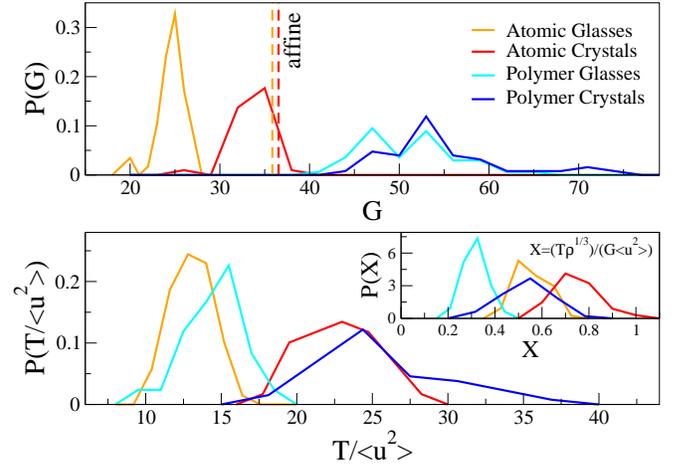}
\end{center}
\caption{Top panel: distributions of the elastic modulus of the samples under study. The vertical dashed lines mark the average affine moduli $G_A$ of the atomic solids
(the corresponding narrow distributions are not plotted for clarity reasons). The affine moduli of the polymeric solids are about one order of magnitude larger, i.e. non-affine effects are much larger.
Modulus is affected by connectivity and, much less, by  crystallinity.
Bottom panel: corresponding distributions of the inverse reduced Debye-Waller (DW) factor, a measure of the local stiffness.
The latter  is poorly affected by connectivity and glasses are locally softer than crystals. Inset: distributions of the quantity $X = T \rho^{1/3}/ G \langle u^2 \rangle$.
Consistency requires that, if the changes of the reduced DW from crystals to glasses are driven by the corresponding elastic changes, their $X$ values must be equal, possibly depending on the kind of solid,
atomic or polymeric, only.}
\label{shear}
\end{figure}

\textbf{\textit{Elasticity and vibrational dynamics.-}}
We perform simple shear deformations of the solids using the Athermal Quasi-Static (AQS) protocol  \cite{BarratAQS13,NicolaElastico}.
An infinitesimal strain increment $\Delta\varepsilon=10^{-5}$  is applied to the simulation box of side $L$ containing the sample, after which the system is allowed to relax in the nearest 
local energy minimum with a steepest descent minimization algorithm. 
Simple shear is performed independently in the planes ($xy$, $xz$, $yz$), and at each strain step in the plane $\alpha\beta$ the corresponding component of the macroscopic stress tensor
$\tau_{\alpha, \beta}$ is taken as the average value of the per-monomer stress $\tau_{\alpha, \beta}^i$ where \cite{allentildesley}:
\begin{equation}
\label{stresstensor2}
\tau_{\alpha, \beta}^i =  \frac{1}{2 \, v} \sum_{j \ne i} 
\frac{r_{\alpha ij} r_{\beta ij}}{r_{ij}} 
\frac{\partial U(r_{ij})}{ \partial r_{ij}}  , \hspace{1cm} \alpha \neq \beta
\end{equation}
where $r_{\gamma  i j}$  is the $\gamma$ component of the position of the $i$ particle with respect to the $j$ particle, $r_{ij}$ is the distance between these two particles,
$U(r_{ij})$ their pairwise additive central potential energy function and $v$ is the average per-monomer volume, $v = L^3/N$. 
The shear elastic modulus $G_{\alpha, \beta}$ is measured as the slope of the resulting stress-strain curve in the linear regime, $\varepsilon \le \varepsilon_{th}= 10^{-4}$, 
as shown in the main panel of Fig.\ref{carico}. An average for each sample is then taken as $G = 1/3 (G_{x,y}+G_{x,z}+G_{y,z})$.
The affine component of the elastic modulus $G_A$ can also be measured with the same procedure, but deforming the simulation box without letting the system relax into a
local energy minimum after each deformation step \cite{BarratAQS13}. The inset of Fig.\ref{carico} shows the distributions of the stress changes per particle $\Delta \tau^i_{\alpha, \beta}$
at the deformation $\varepsilon = \varepsilon_{th}$ following the AQS protocol. Even if the macroscopic response appears to be elastic, plastic events, a few percent,
are signaled by the weak wing of the distribution in the region of negative drops. {\it This is evidence of microscopic elastoplastic heterogeneity preventing an interpretation in terms of a 
homogeneous elastic continuum}. Plasticity of polymeric glasses in the elastic regime has already been evidenced \cite{DePabloBarratPlasticityElasticPolymeGlassPRE08}. Elastic heterogeneity in glasses, is widely reported, e.g. see
\cite{BarratBosonPeakNonAffinePRB02,DePablo04,BarratPRE09}.

Fig.\ref{shear} (top) plots the distributions of the shear moduli of the solids under study. {It is seen that the connectivity increases the rigidity, while crystallinity has weaker influence. In comparison with atomic solids, polymer solids exhibit wider distributions of the modulus, due to the large
influence of the specific chain conformations frozen in the sample}. One notices that not only the modulus, but also its affine $G_A$ and non-affine $G_{NA}$ contributions are larger in polymeric solids with respect to the atomic counterpart.  This is due mainly to the presence of the stiff bonding interactions.

\begin{figure}[t]
\begin{center}
\includegraphics[width= 0.98 \linewidth]{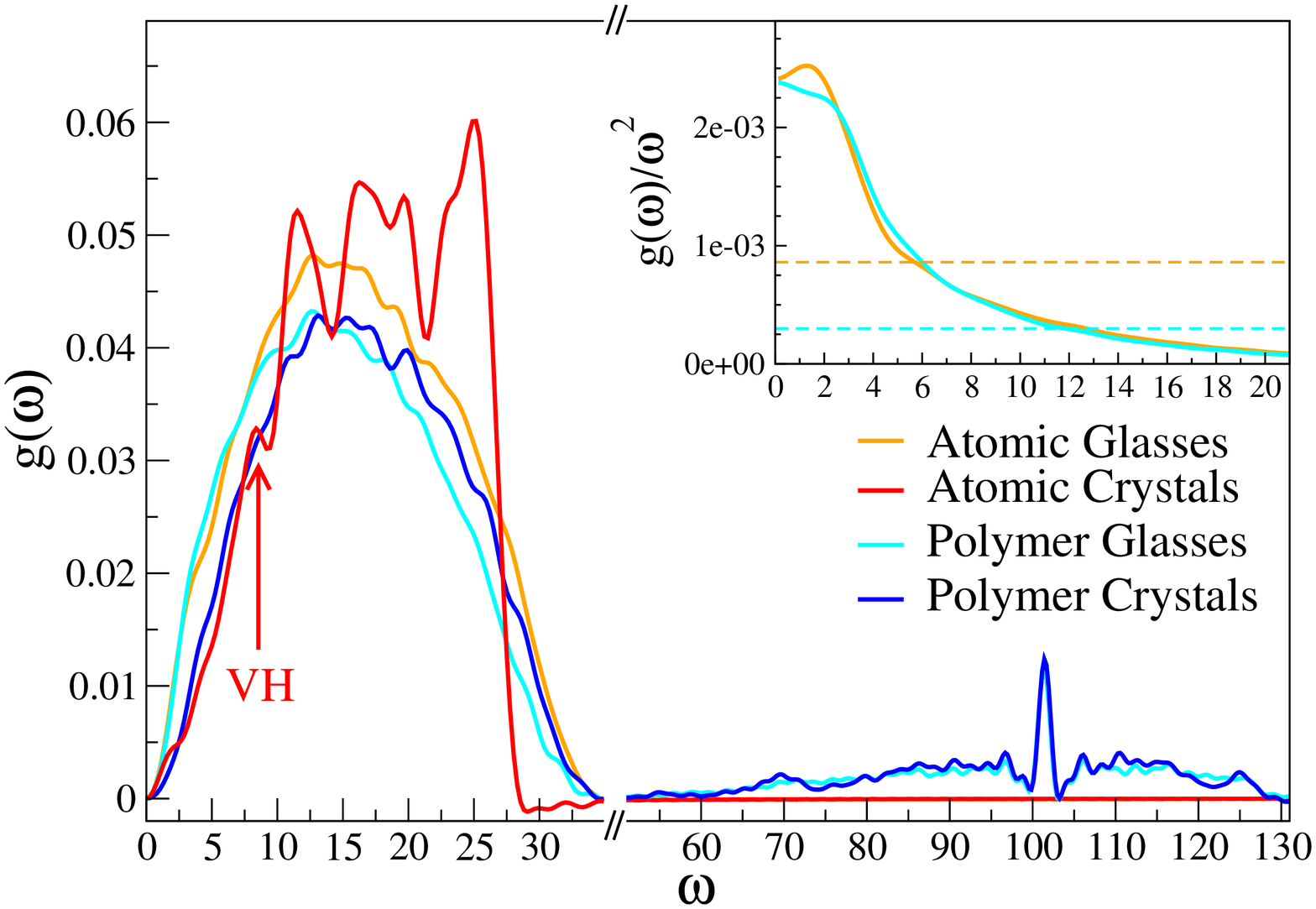}
\end{center}
\caption{Average density of the vibrational states (DOS) of the solids under study. Polymeric solids exhibit a high-frequency structure centered at about $\omega \sim 100$ with about $13 \%$
of all the states \cite{DePabloBPConfinedPolymerJCP04,ZacconeBosonPeakPolymerMM18}. In spite of that, the low-frequency side of polymeric and atomic DOS are nearly coinciding and glasses exhibit an excess
number of states with respect to crystals. The position of the low-frequency van Hove singularity of the bcc atomic crystal is indicated \cite{HafnerDOSBCC}. Inset: reduced DOS of the glasses in the BP region.
The different Debye levels of the two glasses  are marked as horizontal  dashed lines.}
\label{dos}
\end{figure}

We compare the macroscopic elasticity to the local stiffness assessed by the inverse reduced Debye-Waller (DW) factor $T/ \langle u^2 \rangle$,
where the DW $\langle u^2 \rangle$ \cite{Ashcroft76} is taken as the plateau exhibited by the mean square displacement of the particle after the ballistic regime.
We explicitly checked that the reduced DW  is temperature-independent, i.e. unaffected by anharmonicity, up to $T \sim 0.1$. In the harmonic regime DW and DOS are related via \cite{Ashcroft76}:
\begin{equation}
\label{DW}
\frac{\langle u^2 \rangle}{T} \propto  \int_0^\infty \frac{g(\omega)}{\omega^2} \, d \omega.
\end{equation}
Eq.\ref{DW} tells us that the DW is a measure of the role of the low-frequency side of DOS. Fig.\ref{shear} (bottom) plots the distributions of the inverse reduced DW and shows that glasses
are softer than crystals, implying an excess of soft modes in the glass according to Eq.\ref{DW}, and that {\it local softness is poorly affected by connectivity}. 
{\it The finding is in striking contrast with respect to the shear modulus where 
connectivity plays a major role}, see Fig.\ref{shear} (top). According to Eq.\ref{DW}, the negligible influence of connectivity on local stiffness suggests similarity of the low-frequency portion of the DOS
of the atomic and the polymeric glasses.  The results presented in the top and the bottom panels of Fig.\ref{shear} also suggest that the changes of the reduced DW and, by reflection, of the low-frequency
part of DOS from the crystal to the glass are {\it not} accounted for by the corresponding changes of the macroscopic elasticity. 
For that to be true, in fact, consistency with the Debye model \cite{Ashcroft76} would require that the quantity
$X = T \rho^{1/3}/ G \langle u^2 \rangle$ should be the same in glasses and corresponding crystals.
The inset of Fig.\ref{shear} (bottom) presents the distributions of the quantity $X$.
We observe a small overlap between the distributions of the $X$ quantity of crystals and corresponding glasses for both the atomic and the polymeric systems \cite{note2},
i.e. {\it the coupling between elasticity and DW in crystals and glasses is different}.

To better clarify how this finding reflects in the low-frequency side of DOS, we evaluate the DOS from the velocity correlation function and average over all the samples of a specific kind \cite{Rahman76}.
The results are plotted in Fig.\ref{dos}. Polymeric solids exhibit a high-frequency structure centered at about $\omega \sim 100$ 
\cite{DePabloBPConfinedPolymerJCP04,ZacconeBosonPeakPolymerMM18}. The structure corresponds to a fraction of modes, about $13 \%$, with strong involvement of the stiff bonds of the chain. At low frequency, $\omega\lesssim10$, the polymeric DOS is nearly {\it coinciding} with the one of the atomic DOS for both glasses and crystals and the DOS of glasses exhibits an excess number of states with respect to crystals. Modes with $\omega\lesssim10$ virtually do {\it not} involve stiff bonds in polymeric solids. In view of Eq.\ref{DW}, the low-frequency region of DOS is anticipated to dominate the reduced DW. In fact, we observe full correspondence with the results concerning the reduced DW, see Fig.\ref{shear} (bottom). For polymeric solids the low-frequency fraction of modes with no or limited involvement of the stiff bonds is about $87 \%$ and extends up to $\omega \sim 35$.
This is in very rough agreement with the constraint theory of glasses predicting for the present model that, neglecting the non-bonding interactions, about $66 \%$ of the modes are zero-frequency floppy modes  and switching on the weak non-bonding interactions moves the floppy modes slightly away from zero-frequency \cite{CaiThorpeFloppyPRB89}. 
The virtual coincidence of the excess of soft states with respect to crystal in the atomic and the polymeric glasses is best seen in terms of the reduced DOS $g(\omega)/ \omega^2$, which is shown in the inset of Fig.\ref{dos}. We take this finding as strong evidence that the BP of polymeric glasses is largely dominated by {\it non-bonding} interactions. Macroscopic elasticity plays no apparent role in the BP region, as signaled by the coincidence of the reduced DOS irrespective of the rather different elasticity of the polymeric and the atomic glasses, see fig.\ref{shear} (top). The nearly coincident reduced DOS of the polymeric and the atomic glasses clarifies that the different nonaffine moduli $G_{NA}$ of the two glasses shown by Fig.\ref{shear}a follow mainly from the additional contributions by the bonding interactions to the weighting factor $\Gamma_G(\omega)$ in Eq.\ref{GLemMal}. Pressure studies support the conclusion that the decoupling between BP and elasticity exhibits common features in linear polymers  \cite{SokolovSimionescoBPPolymerPRB08}, irrespective of specific aspects like the  molecular-weight dependence of the modulus \cite{SokolovBPpolymJCP97}. This suggests that the dominance of non-bonding interactions explains the decoupling in the whole class of linear polymers. 

Our results point out the role of soft modes in the decoupling between elasticity
and BP. Higher connectivity, beyond the one of linear polymers, reduces
the fraction of soft modes \cite{CaiThorpeFloppyPRB89} and one expects better coupling. Indeed, no decoupling is observed by both simulations \cite{ShengDOSSpringScience91} and experiments \cite{MonacoBPElasticPRL06,FontanaCaponiBPElasticitySilicaPRB07,MonacoFiorettoBP_polymerResinPRL09,ChumakovBPVanHovePRL11} in highly connected systems.

In conclusion, we showed that a class of systems with low or no connectivity exhibits poor evidence that the BP is coupled to elasticity and,
especially, that the latter accounts for the evolution of the BP from the crystalline to the glassy state. In particular,  for polymeric glasses BP and elasticity decouple because BP - differently from elasticity - is controlled by non-bonding interactions only. Our results suggest that the BP in systems with high and low
connectivity is different in nature.

We thank Antonio Tripodo and Francesco Puosi for helpful discussions. 
A generous grant of computing time from IT Center, University of Pisa and Dell${}^\circledR$ Italia is gratefully acknowledged.


\begin{thebibliography}{47}
\bibitem[{not({\natexlab{a}})}]{note}
\bibinfo{howpublished}{Some atomic liquids crystallized with face-centered
  cubic and hexagonal close packed lattices. {T}hey were discarded.}
\bibitem[{not({\natexlab{b}})}]{note2}
\bibinfo{howpublished}{To be more quantitative, we evaluate the joint
  probability $\Pi$ of measuring the same $X$ value from two measures, one from
  the $P_g(X)$ distribution for glasses and one from the $P_c(X)$ distribution
  for the corresponding crystal. $\Pi = \sum_i P_g(X_i) P_c(X_i) ( \Delta X)^2$
  where the sum extends to all the bins sampling the distributions and $\Delta
  X$ is the bin width (of the order of the uncertainty on the single measure).
  We find $\Pi$ about $7 \%$ and $5\%$ for atoms and polymers, respectively}.
\end{thebibliography}
\end{document}